\documentclass{article}

\begin{document}

{\bf STARDRIVES AND SPINOZA }

\bigskip

By: LOUIS CRANE, KSU MATH DEPARTMENT, FQXi GRANT BG0522,

\bigskip

{\bf I. PHYSICS AND LIMITS}

\bigskip

There are several senses in which one may speak of limits in relation
to Physics. 

For the purposes of this essay, it is most useful to begin with the
technological limitations of practical human
activity.  In this context, Physics both provides positive and
negative information. Physical laws tell us how to achieve some goals,
and informs us that others are forever out of reach. 

An excellent historical example is the science of thermodynamics. On
the one hand, it tells us how to construct heat engines, on the other,
it provides us with the Carnotic limit on efficiency, and rules out
perpetual motion machines of both kinds.

So Physics both extends our limits and helps us to understand them. 
Despite the annoying tendency of popular authors to tell us that
anything is possible, the laws of Physics act as negative limits, and
this is one of the great services of science to humanity.

One may also ask about the limits of our knowledge, i.e. of physical
theory. Is it almost finished, as Feynman insisted, does it end in a
non-predictive theory, as we sometimes hear from string theorists, or
is there a final theory yet to be discovered?  

Finally, we can discuss the limits of Science in relationship
to other intellectual disciplines, such as Philosophy and Theology.

In this essay, we shall present possible end results of Physics in
the different senses we discussed above: our practical
possibilities, our understanding of nature, and
the role of Physics in our intellectual and (for want of a better
phrase) spiritual life; and show that they are
closely interwoven.

 We shall outline this network of ideas here, then discuss them
in detail in the rest of the essay.

We begin by discussing a family of potential
practical applications of Theoretical Physics, namely the creation of
artificial black holes (ABH's), and their use as powerplants and
stardrives \cite{CW}. This is suggested by classical general relativity and
quantum field theory, but in order to see if it is truly possible, we
will need a quantum theory of gravity, and moreover, one which can be
put in a certain form, as we shall discuss below. 

Black holes, according to Hawking, are not really black. They radiate
a thermal bath because of quantum field effects on the boundary
\cite{Frolov}. For astronomical black holes the power is minuscule,
but for microscopic ones it can be enormous. We have investigated
whether there is a size range which is both useful and possible to
produce, and found that the answer is a qualified yes.

This proposal is extremely difficult technically.
However, it seems that it is actually the least difficult possible
approach to transporting human beings to the stars. Of the
other suggestions, some are vastly larger and more difficult, others 
seem to be excluded by physical laws much like perpetual motion
machines. 

The ABH proposal seems
just at the edge of possibility. There are physical effects we have
not carefully considered yet which seem likely to make it less difficult.

An ABH embedded in a power plant and transported to a near earth orbit
could provide a permanent solution to our energy problems, provided
ABH's can be fed efficiently. Even if it took a few centuries to
implement the ABH proposal it would be highly desirable for this
reason, as stocks of coal and uranium will run out sooner or later. An
ABH is the ultimate renewable energy source.

Because of the importance of quantum corrections,
this potential application of Physics, which would extend the
range of human activity to interstellar space, is connected to
discovering a new level of physical theory, which very likely will be
the final one.  
 
So the last key opens the last door.

The creation of ABH's would also be the ultimate physical experiment,
in that careful examination of the Hawking radiation as the ABH
evaporated and heated, would reveal to
us all the fundamental particles and forces all the way to the Planck
scale. Thus it would allow us to reach the limit of knowledge of
matter.

Lastly, we shall discuss issues of the role of intelligence in the
universe, and the question of ultimate purpose as applied to humanity
at large. These questions have generally been regarded as beyond the
sphere of science; belonging either to Philosophy and Theology, or to
the category of pseudoquestions, better not pursued at all.

We shall argue that the human production of ABH's, opens the
possibility 
of an approach to these questions from
Physics itself, what we have referred to elsewhere as the
meduso-anthropic principle  \cite{Crane}. Thus, in reaching its terminus, Physics
will transcend its current bounds in the sphere of human thought.

The matters under discussion here are extrapolations beyond the bounds of what is
currently known. They can only have the status of a plausible forecast
whose substantiation or refutation will take a long time.  

\bigskip

{\bf II. THE PATH TO THE STARS? ON THE FEASIBILITY OF ARTIFICIAL BLACK HOLE APPLICATIONS}

\bigskip
In this section we want to discuss whether the Physics of black holes,
together with the laws of Physics of matter as we
know them, make it possible to produce artificial BHs which would be
useful, either as power plants or as starships.

Since the mass of a black hole decreases with its radius, while its
energy output increases and its life expectancy decreases (see \cite{Frolov}), this is a
delicate question.

\
\\
{\bf List of criteria:}
{\it We need a black hole which}
\bigskip
\
\\
{\it 1. has a long enough lifespan to be useful,}
\bigskip
\
\\
{\it 2. is powerful enough to accelerate itself up to a reasonable fraction of the speed of light in a reasonable amount of time,}
\bigskip
\
\\
{\it 3. is small enough that we can access the energy to make it,}

\
\\
{\it 4. is large enough that we can focus the energy to make it,}
\bigskip
\
\\
{\it 5. has mass comparable to a starship,or is smaller but can be fed
  very efficiently.}

\bigskip

We could easily imagine that this would be impossible. Somewhat
surprisingly, it turns out that there is a range of BH radii, which
according to the semiclassical approximation, fit these
criteria.

Using the formulae discussed in \cite{CW} or \cite{Frolov}, we find that a black hole with a radius of a few 
 attometers at least roughly meets the list of
criteria. Such BHs would have mass of the order of 1,000,000 tonnes,
and  lifetimes ranging from decades to centuries. A  high-efficiency square solar panel a few hundred km on each side, in a circular orbit about the sun at a distance of 1,000,000 km,  
would  absorb enough energy in a year to produce one such BH. 

A BH with a life span on the order of a century would emit 
enough energy to accelerate itself to relativistic velocity in a
period of decades. If we could let it get smaller and hotter before feeding 
matter into it, we could get a better performance.

In Section III below, we discuss the plausibility of creating ABHs with a very large spherically converging gamma ray laser.  A radius of 1 attometer corresponds to the wavelength of a  gamma
ray with an energy of about   1.24 TeV. Since the wavelength of the Hawking
radiation is $8\pi^2$ times the radius of the BH, the Hawking temperature of a BH with this radius  is on the order of 16 GeV, 
within the limit of what we could hope
to achieve in a nuclear laser.

Now the idea that the wavelength of the radiation should match the
radius of the BH created is very likely pessimistic. The collapsing
sphere of radiation would gain energy from its self-gravitation as it
converged, and there is likely to be a gravitational self-focussing.

Thus it seems that making an artificial black hole and using it to
drive a starship is just possible, because the family of BH solutions
has a ``sweet spot.'' This has not been remarked before in
the literature, except for the earlier work of the author \cite{Crane}.

We cannot fully trust this result, in that we do not know how to take
corrections from quantum gravity into account.

\bigskip

{\bf III. FOUR MACHINES}

\bigskip

Now we discuss how our proposal could be implemented technically. 
The devices we propose are far beyond current technology, but we
do not see why they would be impossible if an advanced future industrial society
were determined to make them.

\bigskip

{\bf A. The black hole generator}
\bigskip

In a previous paper by the author \cite{Crane}, it was proposed that an ABH 
could be artificially  created by firing a huge number of  gamma rays
from a spherically converging laser. The idea is to pack so much
energy into such a small space that a  BH will form.  An advantage of
using photons is that, since they are bosons,  there is no Pauli
exclusion principle to worry about. Although a laser-powered black
hole generator presents huge engineering challenges, the  concept is physically sound according to classical general relativity. The Vaidya-Papapetrou metric shows  that an imploding spherically symmetric shell of ``null dust" can form a black hole  (see, e.g., \cite{Frolov}, p. 187, or Joshi \cite{Joshi} for further details).

Since photons  have null stress energy just like null dust,
 a black hole should form if a large aggregate of photons interacts classically with the gravitational
field. As long as we are discussing regions of spacetime that are many
orders of  magnitude larger than the Planck length, we should be
outside of the regime of quantum gravity and classical theory should
be appropriate.

Since a nuclear laser can convert on the order of $10^{-3}$ of its
rest mass to radiation, we would need a lasing mass of order $10^9$
tonnes to produce the pulse. This should correspond to a mass of order
$10^{10}$ tonnes for the whole structure (the size of a small asteroid). Such a structure would be
assembled in space near the sun by an army of robots and built out of
space-based materials. It is not enormously larger than some structures human
beings have already built. The precision required to focus the
collapsing electromagnetic wave would be of an order already possible
using interferometric methods, but on a truly massive scale.

This is clearly extremely ambitious, but we do not see why it should
be impossible. Given the enormous advantages it could bring to
humanity, it should be approached as optimistically as the facts permit. 

\bigskip

{\bf B. The drive} 

\bigskip
Now we would like to discuss how to use an ABH to drive a starship. We
need to accomplish 3 things.

\bigskip

{\bf Design requirements for a BH starship}

\bigskip

{\it 1. use the Hawking radiation to drive the vessel}

\bigskip

{\it 2. drive the BH at the same acceleration}

\bigskip

{\it 3. feed the BH to maintain its temperature}

\bigskip

Item 3 is not absolutely necessary. We could manufacture an ABH, use it
to drive a ship one way, and release the remnant at the
destination. However this would limit us greatly as to performance,
and be very disappointing in the powerplant application discussed below.

We shall discuss these three problems in outline only here; at the
level of engineering they will require years of study.

It is not hard to see how to approach requirement 1. We 
position the ABH at the focus of a parabolic reflector attached to the
body of the ship. Since the ABH will radiate  gamma rays and a mix of
particles and antiparticles, it is not clear how to construct the
reflector. 
It has already been
proposed
to make a gamma ray reflector
out of an electron gas  in the closely related case of an antimatter
rocket \cite{Sanger}.

It is not clear if this is feasible (e.g., \cite{Forward}).

A more detailed discussion of the drive design appears in \cite{CW}.

The most optimistic approach is to solve requirements 2 and 3 together
by attaching particle beams to the body of the ship behind the ABH and
beaming in matter. This would both accelerate the ABH, since BHs ``move when
you push them''(see  \cite{Frolov} p270), 
and add mass to the ABH, extending its lifetime.

The delicate thing here is the absorption cross section for a particle
going into a BH. If simply
aiming the beam at the ABH doesn't work, we can try forming an
accretion disk near the ABH and rely on particles to tunnel into
it. Alternatively, we could use a small cluster of ABHs instead of just
one to create a larger effective target, charge the ABH etc. It is
also possible that because of quantum effects ABHs have larger than
classical radii, due to the analog of zero point energy. 

There are
some physical considerations which we are still investigating, which
make us optimistic on this point. Nevertheless it must remain as a 
challenge for the future.
\bigskip

{\bf C. The powerplant}
\bigskip

This has already been proposed by Hawking (see \cite{BHOT} pp. 108 - 109). We simply surround the ABH with a spherical shield, and use it to
drive heat engines. (Or possibly use gamma ray solar cells, if such
things be.) This would have an enormous advantage over solar electric
power in that the energy would be dense and hence cheaper to accumulate.

The 3 machines here really form a tool set. Without the drive, getting
the powerplant near Earth where we need it would be very
difficult. Without the generator, it would require the good fortune to
find a primordial ABH to implement the proposal.

\bigskip

{\bf D. The self-driven generator}

\bigskip 
The industry formed by our first 3 machines would not yet be really
mature. To fully tap the possibilities we would need a fourth machine,
a generator coupled to a family of ABHs which could be used to charge
its laser. Assuming we can feed a ABH as discussed above, we would
then have a perpetual source of ABHs, which could run indefinitely on
water or dust or whatever matter was most convenient.

A civilization  equipped with our four machine tool set would be almost
unimaginably energy rich. It could settle the galaxy at will.

\bigskip

{\bf IV.THE ONLY PATH?}

\bigskip

The technical program we have
outlined is extremely difficult. Why even consider it? Because the
goal of interstellar flight is so profoundly important to our future,
and because the alternatives are either much more difficult and much
larger, or actually impossible barring a major surprise in
Physics. 

Also, a long term solution to our energy problems is not yet
clearly at hand. ABH technology would provide us with energy resources
beyond even fusion power, and not use anything except waste matter.

\bigskip

{\bf A. Shielding}

\bigskip

The dreams of manned spaceflight of the fifties and sixties have
largely gone unrealized. This is to an important degree because it has been
discovered that cosmic radiation has much more serious medical 
consequences than was originally believed \cite{SciAm}.

 Substantial human presence
in space has only occurred in low Earth orbit near the equator, where
the earth's magnetic field shields us from the cosmic radiation.

Our visits to the moon were brief, and it is known that a
prolonged human presence there would have to be underground,
necessitating the transport of massive ``earth'' moving equipment
\cite{SciAm}.

It is now known that 
any prolonged human presence deeper in space would need to be behind a shield
of the effective strength of two feet of lead, which would weigh 400
tonnes for a small capsule \cite{SciAm}.

It therefore becomes more economical to
think of a larger vessel, weighing many thousands of tonnes, in which
a group of people could live indefinitely. This possibility has not 
been very widely explored, particularly not in a practical direction,
because of the enormous energies involved
in accelerating such a body. 
\bigskip

{\bf B. Specific impulse} 

\bigskip

The distances between the stars are so great that practical travel between them
would require us to reach speeds comparable to the speed of
light. This is extremely difficult to do because very few processes
known to us release energies comparable to the masses of the matter
involved. Nuclear reactions, for example, release only a fraction of a
percent of the rest masses of the nucleii, so that an interstellar
vehicle powered by fission or fusion would have to carry many 
thousands of times the mass of its payload in fuel.

Coupled with the
shielding problem discussed in Section A above, this means an
interstellar voyage using nuclear energy would deplete the Earth's
resources of fissile or fusile materials to an intolerable degree.

Other than black hole radiation the only
process we know of which is sufficiently energetic for interstellar
flight is
matter-antimatter annihilation. 
This has been proposed, but there are two
severe obstacles. 

The first is that the efficiency of antimatter production
in current accelerators is well below $10^{-7}$ (very few collisions produce a
trappable antiparticle) \cite{Forward}. Thus, making enough antimatter to propel a
starship would use up ten million times as much energy as our
proposal. The most optimistic projections of antimatter enthusiasts do
not produce an efficiency above $10^{-4}$, so that at best our
proposal is still ten thousand times more efficient.

The second obstacle is containment. A microscopic particle of
ordinary matter which drifted into the antimatter would cause an
explosion, scattering the antimatter into contact with the ship, and
destroying everything for millions of miles around. Any
electromagnetic force which held the antimatter in would also drive
normal matter in. One hears the suggestion that this could be solved by
``magnetic bottles,'' but magnetism is a force which acts
perpendicularly to the motion of a charged particle, and therefore
does not in any simple way form a bottle. Experiments in the  magnetic
confinement of plasma for millisecond intervals have been very frustrating.

Paramagnetic forces can repel matter, but they are extremely weak, and
treat matter and antimatter identically. So they would force normal
matter in as much as antimatter.

For these reasons an antimatter starship seems out of reach given
Physics as we know it. We could imagine surprises in future Physics
which would change this picture, but they seem remote. Dark
matter, for example, interacts neither with normal matter nor
antimatter.

Let us briefly contrast the idea of using antimatter with the
ABH proposal. It is currently possible
to produce antimatter in extremely small quantities, while a synthetic
black hole would necessarily be very massive. On the other hand, the
process of generating an ABH from collapse is naturally efficient, so it
would require millions of times less energy than a comparable amount
of antimatter or at least tens of thousands of times given some
optimistic future antimatter generator. 

As to confinement, a BH confines itself. We would need to avoid
colliding with it or losing it, but it won't explode. Matter striking a
BH would fall into it and add to its mass. So making a BH is extremely
difficult, but it would not be as dangerous or hard to handle as a
massive quantity of antimatter. 

Although the process of generating an
ABH is extremely massive, it does not require any new Physics. Also, if an
ABH, once created, absorbs new matter, it will radiate it, thus acting
as a new energy source; while antimatter can only act as a storage
mechanism for energy which has been collected elsewhere and converted
at extremely low efficiency.

None of the
other ideas suggested for interstellar flight seem viable either.
The proposal for an interstellar ramjet turns out to produce more drag
than thrust \cite{new}, while the idea of propelling a ship with a laser beam
runs into the problem that the beam spreads too fast.

At this point, we do not even have a viable method for
sending very small probes to other stars. Sending human beings is much
harder, but in some sense it is what we really want, being what we are.

\bigskip

{\bf C. Wormholes and Warpdrives}

\bigskip

The mere fact that wormholes or warpdrives are discussed in the
context of interstellar flight shows how difficult a goal it is, and
how strongly we desire it.

Both suggestions are forbidden by the positive energy condition. It is
true that it is possible to produce negative energy densities between
the plates of a capacitor via the Casimir effect, but they are
extremely small, and could never remotely equal the masses of the
plates. Starting from positive energy matter, which is all we have
ever seen, there is no method anyone knows to produce the enormous
negative energy densities a warp drive or wormhole would require. 

It is by far the most probable case that they are simply impossible.

\bigskip

{\bf V. CHALLENGES TO QUANTUM GRAVITY IN THE ABH PROPOSAL}

\bigskip
If a theory of quantum gravity is to allow us to find the necessary
corrections to see if the ABH proposal is feasible, it must allow us
to describe quantum effects in a small highly energetic region whose
surroundings can be treated classically and well approximated by flat
spacetime. A quantum theory of the whole universe would need profound
reinterpretation to be used for this purpose.

This suggests a version of quantum gravity which is defined relative
to an observer, which is treated as classical. It is widely believed
that the information which can pass from a bounded region of spacetime
to an external observer is finite, limited by the Bekenstein
bound \cite{Bekenstein}. Creation of a theory of quantum gravity which includes the
Bekenstein bound in its structure is therefore strongly suggested by
the ABH proposal \cite{Crane2}.

A quantum theory of gravity would also have to describe the
relationship between spacetime and matter, in order to improve on the
semiclassical picture which led us to the prediction of Hawking
radiation. Since Hawking radiation is generated from transplanckian
modes near the horizon of the BH which are then redshifted, it is
highly unlikely that the semiclassical calculations would be accurate
enough for the starship designers of the future. 

Such a theory would have to be a unified field theory, if it were to
fully describe the interactions of quantum gravity with all forms of
matter. In all likelihood, it would be a final theory.

\bigskip

{\bf VI. THE
  MEDUSO-ANTHROPIC PRINCIPLE}

\bigskip

The origin of the ABH proposal is very peculiar. The author was
reviewing the work of Lee Smolin, which was later published in a book 
entitled \emph{The Life of the Cosmos} \cite{Smolin}. 
Professor Smolin proposed that the universe we see was
only one of many universes, and that new universes arise from old ones
whenever a black hole is produced. This then leads to an evolutionary
process for universes in which universes with an unusually high number
of stars are selected for.
 
The author proposed that the evolutionary process of universes
should include life. This is possible if successful industrial
civilizations eventually produce black holes, and therefore baby universes.
                               
Note that if successful industrial civilizations only trapped already
existing BHs, it would not alter the number of baby universes a
universe produced, so that no evolutionary loop would result.

This led us to consider the possibility of producing ABHs,
and to explore how they might be useful.

The meduso-anthropic principle (as this proposal was named) is at
least falsifiable, in the sense that if ABHs
prove to be completely  impossible or useless then the evolutionary cycle
universes-civilizations-black holes -baby universes could never happen.

The result of our feasibility calculation is much too tenuous to be
considered evidence for the meduso-anthropic principle. It is not in
fact certain that black holes create baby universes, although the
maximal analytic continuations of the standard BH solutions to
Einstein's equation suggest that they do. 

Nevertheless, only through this line of thought did we
consider the possibility of synthetic BH creation. We are the first
and almost the
only author to our knowledge to consider this. The only other author we
are aware of is Semiz \cite{Semiz}, who wrote: ``...we would have to
either find small black holes, possibly primordial, or manufacture
them by means as yet unknown,'' in a rather popular discussion on BH
powerplants and stardrives, which appeared after our first paper.

\bigskip

{\bf VII. THE ROLE OF SCIENCE IN HUMAN LIFE.  SCIENCE AND RELIGION}

\bigskip
The construction of ABH technology would be preceded by a long period
in which space based industries were constructed and expanded. Armies
of sophisticated automatic machines would be assembled, sources of raw
materials hunted for, and laser technologies advanced.
 
A society that decided to build the laser and other machinery to
implement the ABH proposal would have to allocate resources comparable
to building the pyramids over a timescale comparable to building the
cathedrals. The work would have its fruition only long after the initial
builders' lifespans were over.

If the meduso-anthropic hypothesis turns out to be valid, the builders
would understand their work as part of the eternal recreation of the
universe, having purpose in the sense that the organs of animals
develop purpose as the result of an evolutionary process. Their entire
lives would have a higher purpose in that they result in the creation
of new universes and new life, as well as spreading their descendants
throughout the universe. It would also imply that our own universe is
fine tuned because it is the result of the activity of earlier
intelligence. 

It is impossible to consider these things and not be reminded of
religious ideas. Spinoza's pantheistic idea of God as identical to the
creative power of the universe would take on a much richer
interpretation in such a future.  Christian ideas about the unity of
God and man, and Buddhist or Hindu ideas about the eternal return could find a
home within Science in such a future;  although in a very new sense.

So possibly the quantum theory of gravity will give us the tools to reach the
stars, solve our energy problem forever, teach us we really have
purpose in the 
cosmos, and compel us
to organize ourselves around a common purpose which extends beyond
ourselves. 
Our limits in space and
time will be expanded as the structure of our social consciousness
transforms and unifies itself. Religion will fuse into Science.

Many observers have felt that Western civilisation is experiencing a
time of aimlessness and confusion as its traditional spiritual ideas
fall into apparently hopeless contradiction with the discoveries of
modern Science. Deep social and emotional needs seem irreconcilable
with hard, undeniable facts.

We have outlined a plausible future where a central project for the whole world
gives us an almost infinite extension of our range as a result of an immense
collective effort. Depending on subtle points of interpretation of
general relativity, this could come with a new understanding of our
place in the universe, in which we play a role in its cycle of
creation.

Will this mean that the time of alienation and aimlessness will end,
if we simply have the courage to persevere? Will we find a different
sense of meaning and belonging in the vast group effort to settle the
universe, recreating the multiiverse itself in the effort?

We do not really know. I can say that for myself no other focus than
research into quantum gravity is thinkable any more.

\bigskip

{\bf ACKNOWLEDGEMENTS: } I would never have found the time to write
this without my grant from FQXi. My student Shawn Westmoreland did a
lot of the calculations on which this essay is based, and made helpful
comments.

\end{document}